\documentstyle[twocolumn,eclepsf]{jpsj}

\def\runtitle{Charge Ordering and Spin Gap in 
NaV$_2$O$_5$}
\def\runauthor{Hitoshi {\sc Seo} and Hidetoshi {\sc Fukuyama} }

\def\lsim{\lower -0.3ex \hbox{$<$} \kern -0.75em \lower 0.7ex \hbox{$\sim$}}
\def\gsim{\lower -0.3ex \hbox{$>$} \kern -0.75em \lower 0.7ex \hbox{$\sim$}}

\title{Charge Ordering and Spin Gap in 
NaV$_2$O$_5$}
\author{Hitoshi {\sc Seo}\footnote{E-mail: hseo@watson.phys.s.u-tokyo.ac.jp}
and Hidetoshi {\sc Fukuyama}}
\inst{Department of Physics, Faculty of Science, University of Tokyo, Tokyo 113}

\recdate{}

\abst{A possible ground state of NaV$_2$O$_5$ is proposed
based on the Hartree approximation for both 
on-site and intersite Coulomb interactions. 
The results indicate 
that the intersite Coulomb interaction induces a zigzag type of 
charge disproportionation (i.e. charge ordering) along the ladders of V-ions 
resulting in 
the localized spins between neighboring ladders 
to form a spin gap. 
This new state, which is different from the spin-Peierls state 
so far believed, seems to be consistent with the 
existing experimental results.}
\kword{NaV$_2$O$_5$, spin gap, charge ordering, 
Hartree approximation, spin-Peierls, charge disproportionation}

\begin{document}
\sloppy
\maketitle

Recently NaV$_2$O$_5$ has been attracting interest.\cite{Isobe} 
Based on early structural study which postulated two different V sites, 
i.e. V$^{4+}$-ions and nonmagnetic V$^{5+}$-ions,\cite{Carpy} 
this compound has been considered to be 
described by a simple spin 1/2 Heisenberg chain consisting of V$^{4+}$ 
resulting in the spin-Peierls (SP) state. 

However, there are many controversial issues 
as regards the nature of this material. 
The X-ray superlattice reflection below 
the transition temperature $T_c=35$ K
where the magnetic susceptibility steeply decreases, 
shows that the unit cell gets four times larger along the $c$-axis 
(the chains are along the $b$-axis), which is not 
expected from simple SP dimerized lattice distortion.\cite{Fujii} 
On the other hand, 
in an inelastic neutron scattering measurement below $T_c$, 
two exitation branches along the $a^*$-axis together with 
an intensity modulation were observed along the $a^*$-axis, 
which can not be explained 
based on simple 1D SP chains.\cite{Yoshihama} 
Furthermore, a recent crystal structural analysis showed that 
the V-ions are all equivalent at room temperature
thus have valences 4.5+,\cite{Smolinski,von}
while a $^{51}$V-NMR experiment on single-crystalline sample of 
NaV$_2$O$_5$ revealed that these V sites change its valences at $T_c$ 
from uniform V$^{4.5+}$ to 
two different sites of V$^{4+}$ and V$^{5+}$.\cite{Ohama} 
At the same time, $\Delta/T_{c}$, 
where $\Delta$ is the spin gap energy, is 
significantly large compared to other SP compounds,  
e.g. TTF-CuBDT, MEM-(TCNQ)$_2$ and CuGeO$_3$.\cite{Fujii}

The causes of these puzzling experimental results is poorly understood, 
and even the origin of the observed insulating behavior 
in the whole temperature range is not clear yet. 
As regards the orbitals of $d$-electrons at V sites, 
those with $d_{xy}$ symmetry are suggested to be relevant 
above and below $T_c$,\cite{Ohama2,Smolinski} 
and then this $d_{xy}$-band is nominally 1/4-filled. 
There are theoretical studies based on the assumption of 
charge ordering into chains of 
V$^{4+}$ and V$^{5+}$ on one hand,\cite{Mila,Augier} 
while there is a proposal that 
each electron is located in 
a bonding V-O-V molecular wavefunction.\cite{Smolinski,Horsch}
However, as has been clarified 
in charge-transfer organic compounds,\cite{Kino,Seo}
there exist various possibilities of charge distribution 
in 1/4-filled systems triggered by the intersite Coulomb interaction 
when the geometry and consequently transfer integrals 
of the system are very anisotropic. 

In this letter, 
the possobility of a charge ordering 
in the ground state of NaV$_2$O$_5$ has been pursued by use of the 
Hartree calculations 
including intersite Coulomb interaction. 
Furthermore, the origin of the spin gap behavior is discussed 
based on these results. 

\begin{figure}
\begin{center}
\epsfile{file= 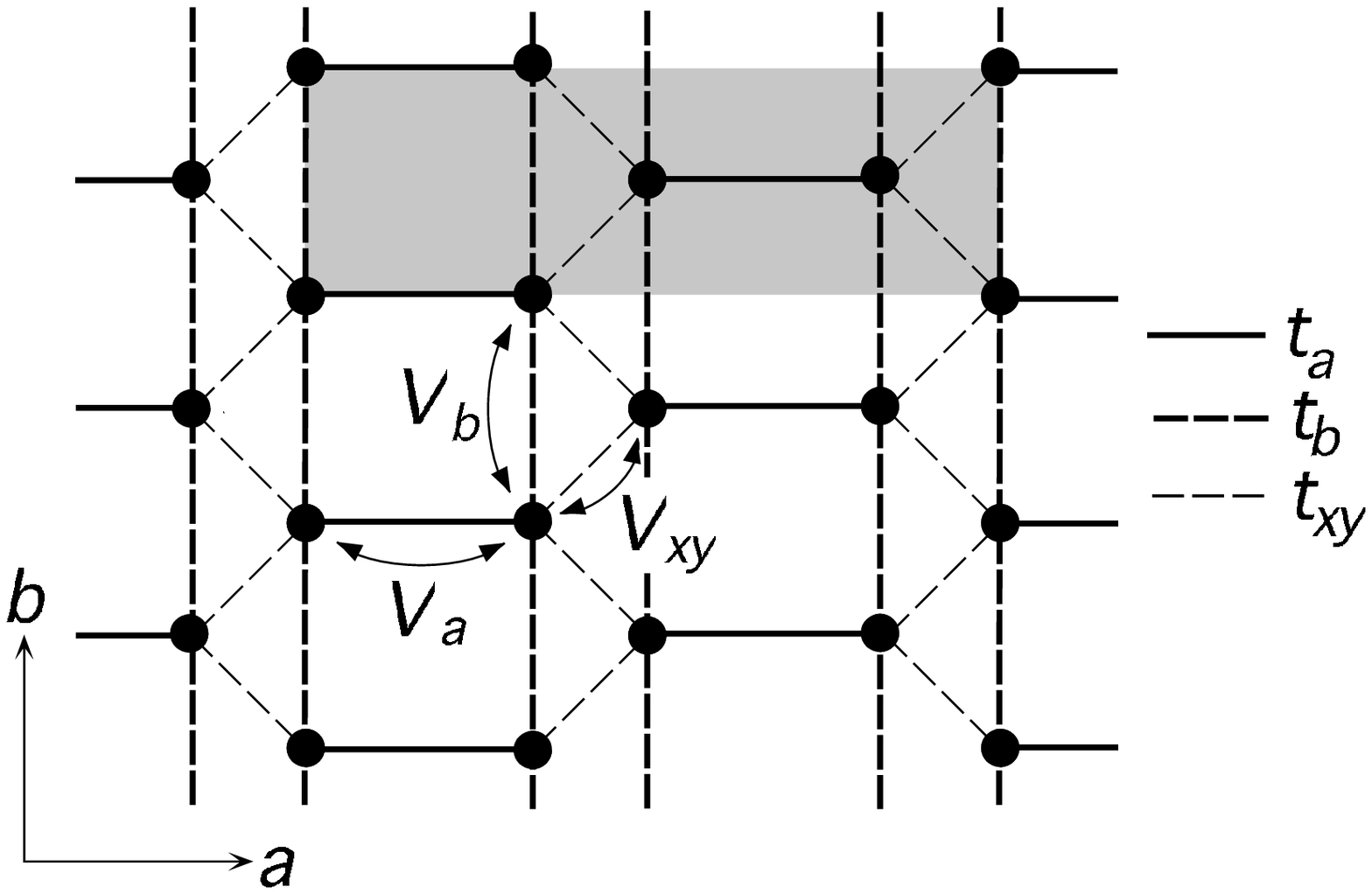,height= 0.5cm}
\end{center}
\caption{A schematic representation of the structure 
of V ions in the $a$-$b$ plane of NaV$_2$O$_5$, 
where V ions are represented as black circles, 
together with the hopping integrals 
(after the notations of ref. 11), 
and the intersite Coulomb interactions 
$V_{i,j}$. Thick, thick-dotted, and thin-dotted lines representing 
the hopping parameters $t_a, t_b$ and $t_{xy}$, respectively. }
\label{structure}
\end{figure}

To investigate the effect of Coulomb interaction, 
the V$_2$O$_5$ layer, namely its $a$-$b$ plane is considered, 
where the spatail arrangement of V-ions 
is schematically shown in Fig. \ref{structure}. 
They form two-leg ladders in the $b$-direction and 
these ladders align in the $a$-direction; 
a unit cell contains four vanadiums as in Fig. \ref{structure}. 
The hopping parameters between them, 
$t_a, t_b, t_{xy}$, the on-site 
and the intersite Coulomb interactions, 
$U$ and $V_{i,j}$, respectively, 
are taken into account; we define $V_{i,j}=V_a,V_b,V_{xy}$ as the 
interaction along the bond with hopping $t_a,t_b,t_{xy}$, respectively, 
as in Fig. \ref{structure}. 
Our Hamiltonian is a model of single orbital, $d_{xy}$, 
for each vanadium site given as follows; 
\begin{eqnarray}
H=\sum_{<i,j>}&\sum_{\sigma}&\left(-t_{i,j}a^{\dagger}_{i\sigma}
    a_{j\sigma}+h.c.\right) \nonumber\\
   &+&\sum_{i} U n_{i\uparrow}n_{i\downarrow}
    +\sum_{<i,j>}V_{i,j} n_in_j,
\label{eqn:Hamil}
\end{eqnarray}
where $<i,j>$ denotes the neighbor site pair, 
$\sigma$ is a spin index which takes $\uparrow$ and $\downarrow$, 
$n_{i\sigma}$ and  $a^{\dagger}_{i\sigma}$ ($a_{i\sigma}$) denote 
the number operator and the creation (annihilation) operator for the 
electron of spin $\sigma$ at the $i$th site, respectively, 
and $n_i=n_{i\uparrow}+n_{j\downarrow}$.  

The Coulomb interactions $U$, $V_{i,j}$ are treated 
within the Hartree approximation 
in a manner similar to that in refs. 12 and 13, 
\begin{eqnarray}
n_{i\uparrow}n_{i\downarrow} &\rightarrow& 
\left<n_{i\uparrow}\right>n_{i\downarrow}
+n_{i\uparrow}\left<n_{i\downarrow}\right>
-\left<n_{i\uparrow}\right>\left<n_{i\downarrow}\right> \nonumber\\
n_in_j &\rightarrow&  \left<n_i\right>n_j
+n_i\left<n_j\right>
- \left<n_i\right>\left<n_j\right>, 
\end{eqnarray}
and self-consistent solutions are searched for. 
Our calculations are carried out at $T=0$ and 
the average electron density is fixed 
at 1/4 per vanadium site. 
Several types of antiferromagnetic (AF) solutions 
are obtained and 
their energies are compared
so that the true ground state is determined. 
The total energy ${\cal E}$ is calculated as 
\begin{eqnarray}
{\cal E}=\frac 1{N_{cell}}&\sum_{lk\sigma}&\epsilon_{lk\sigma}
n_F\left(\epsilon_{lk\sigma}\right) \nonumber\\
-\sum_i&U&\left<n_{i\uparrow}\right>\left<n_{i\downarrow}\right>
-\sum_{<i,j>}V_{i,j} \left<n_i\right>\left<n_j\right>,
\end{eqnarray}
where $N_{cell}$ is the total number of the cells, 
$\epsilon_{lk\sigma}$ is the $l$th eigenvalue at each $k$ and 
$n_F\left(\epsilon \right)$ is the Fermi distribution function. 

\begin{figure}
\begin{center}
\epsfile{file= 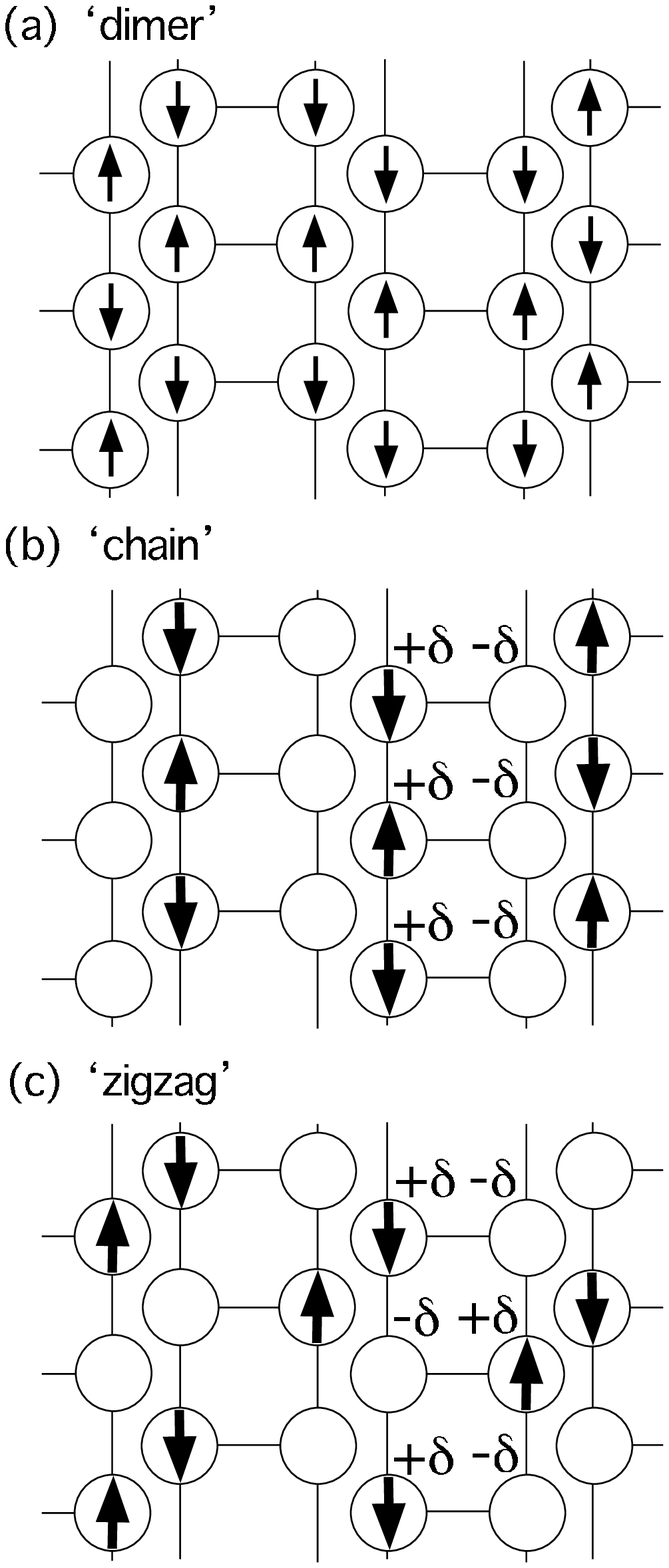,height= 0.5cm}
\end{center}
\caption{A schematic view of 
different types of AF ordering for the structure given in 
Fig. \protect\ref{structure}. Circles and arrows 
represent V-ions and the spin moments, and  
$+\delta$ and $-\delta$ denote the amount of charge disproportionation 
(charge ordering) 
0.5+$\delta$ and 0.5$-\delta$, namely more or less electrons 
than the average, respectively.}
\label{AF}
\end{figure}

The AF solutions we found to be stable are shown in Fig. \ref{AF}.  
The `dimer' type AF (Fig. \ref{AF}(a) ) has uniform charge density, 
i.e. 0.5 electron per site, and, 
if the pairs of vanadium are taken as dimers, 
it can be considered as AF between these dimers along the ladders, 
which corresponds to the state assumed in refs. 5 and 11 
In the `chain' type AF (Fig. \ref{AF}(b)), 
charge disproportionation (charge ordering) exists; 
the charges concentrate on one leg of the ladders, 
and AF ordering emerges along these chains. 
This type corresponds to the conventional 1D V$^{4+}$ spin 1/2 chains.
The `zigzag' type AF is the one newly found, which 
also has the charge disproportionation 
but the sites with more electrons, i.e. V$^{4+}$, 
are distributed in a zigzag way along the ladders 
as in Fig. \ref{AF}(c). 
In the actual solutions obtained by Hartree calculations, 
there exist, to be precise, finite but small spin moments on the sites 
which are described as open circles. 
However, as we will see below, within realistic range of parameters, 
these spin moments are small so that they can be neglected.  

\begin{figure}
\begin{center}
\epsfile{file= 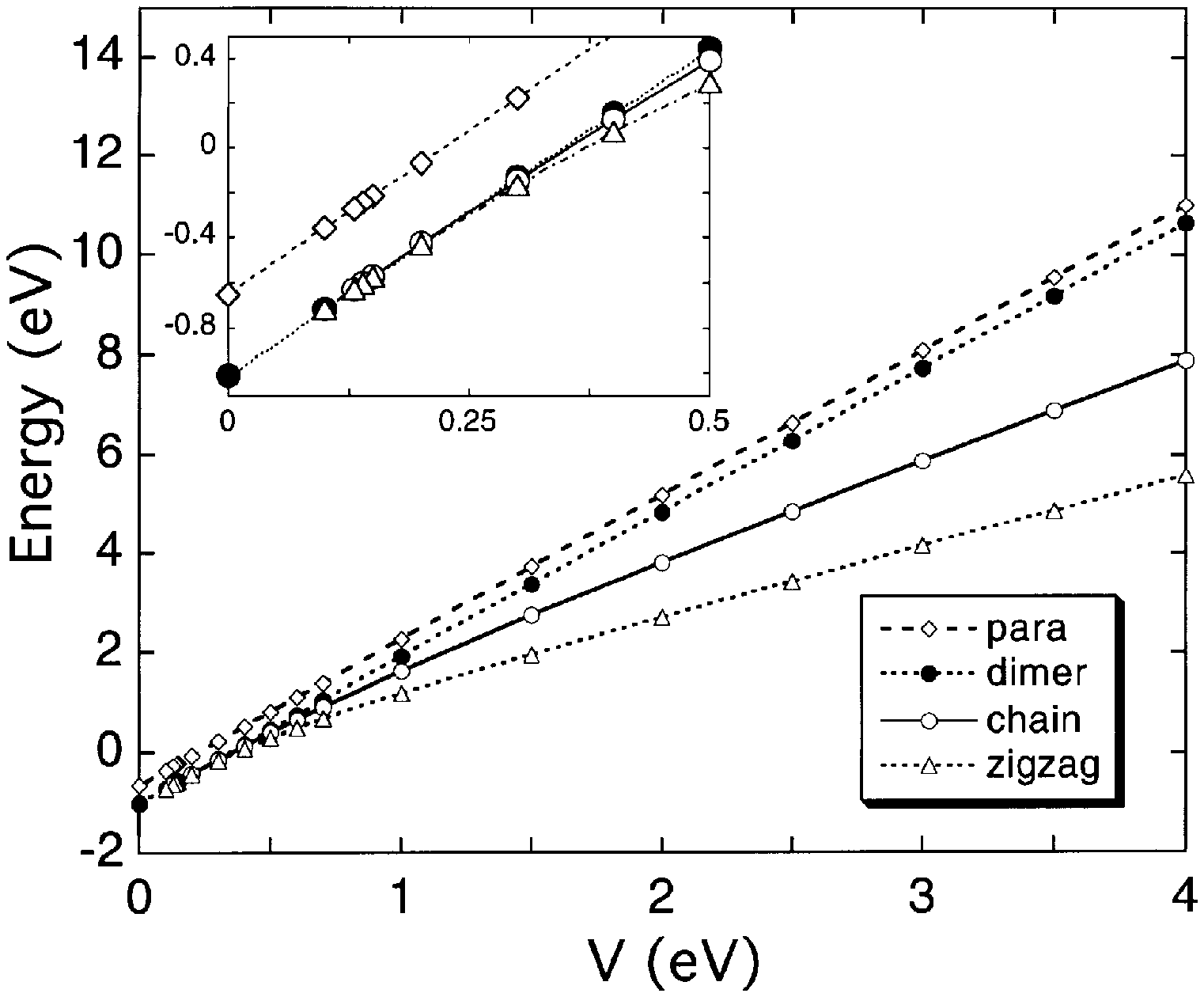,height= 0.5cm}
\end{center}
\caption{The energy per unit magnetic cell of each state shown in Fig. \protect\ref{AF}. 
The inset shows the region of small $V$. }
\label{energy}
\end{figure}

\begin{figure}
\begin{center}
\epsfile{file= 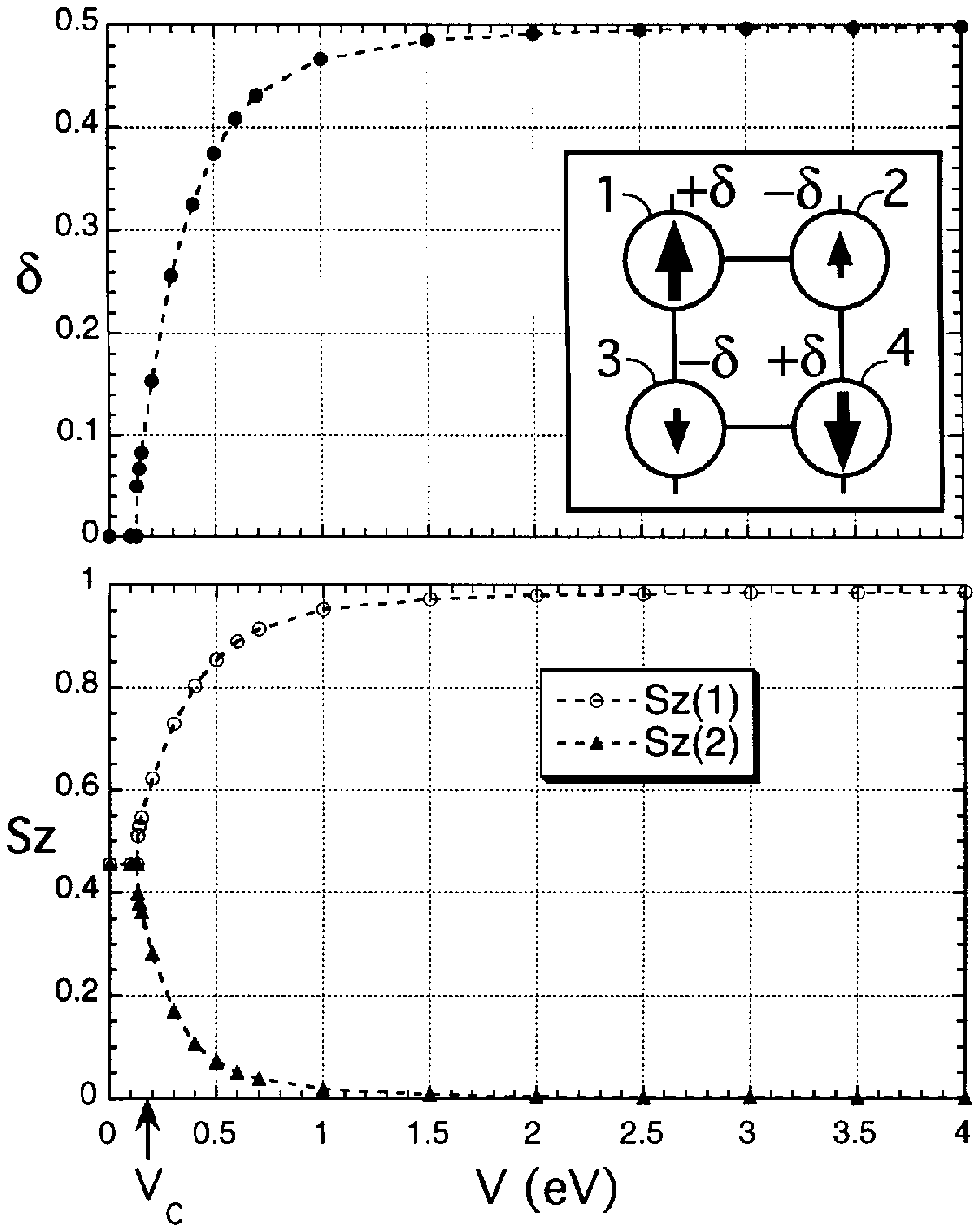,height= 0.5cm}
\end{center}
\caption{The spin moment, S$_z$, and 
the degree of charge disproportionation, $\delta$, 
for the zigzag type AF as a function of V. 
The inset shows the magnetic unit cell along the ladder 
with the site indices, 
where the alignement of the spin moments 
are indicated by arrows and 
the amount of charge is 0.5+$\delta$ and 0.5-$\delta$, respectively.}
\label{Sz}
\end{figure}

For the explicit calculations 
the values of parameters are taken from ref. 11
as $t_a=0.35$ eV, $t_b=0.15$ eV, $t_{xy}=0.3$ eV and $U=4.0$ eV, 
and $V_{i,j}$ are varied maintaining the condition
$V\equiv V_a=V_b=V_{xy}/\sqrt{2}$. 
Here we assumed $V_{i,j}\propto 1/d_{i,j}$,  
where $d_{i,j}$ is the distance between the $<i,j>$ pair, 
and approximated that 
the distances between intraladder vanadiums are all same
and the distance of interladder vanadiums is 
$\sqrt{2}$ times smaller than that. 
The calculated energies of each state  
as a function of $V$ 
for three different types of AF solutions 
together with that for the paramagnetic solution 
are shown in Fig. \ref{energy}.
This result shows that `zigzag' type of AF has the lowest energy
within a range of realistic values of $V$, namely $V\sim 1$ eV. 
The magnitudes of spin moment on sites 1 and 2 
(see the inset of Fig. \ref{Sz} where a unit magnetic cell is shown), 
S$_z$($i$) ($i$=1,2), and 
the amount of charge disproportionation $\delta$
in this case of `zigzag' type AF 
are shown in Fig. \ref{Sz}.
We note that the charge densities are 0.5+$\delta$ on sites 1 and 4, 
0.5$-\delta$ on sites 2 and 3, 
and the spin moments have the relation 
S$_z$(1)=$-$S$_z$(4) and S$_z$(2)=$-$S$_z$(3). 
When $V$ is decreased, 
this `zigzag' state gives away continuously to the `dimer' state at 
$V=V_c$, 
where S$_z$(1)=S$_z$(2)=$-$S$_z$(3)=$-$S$_z$(4). 
When $V\gsim 0.5$ eV, which is expected to be realistic, 
$\delta\simeq$0.5, i.e. the charge is almost concentrated in 
sites 1 and 4, 
and $|$S$_z$(1)$|$=$|$S$_z$(4)$|\simeq$1.0  
$(|$S$_z$(2)$|$=$|$S$_z$(3)$|\simeq0)$. 
Hence we can consider that 
the electrons are almost fully localized on sites 1 and 4 
to form V$^{4+}$ while V$^{5+}$ in sites 2 and 3. 
If quantum fluctuation is taken into account 
in this state where localized 1/2 spins exist on the sites 1 and 4, 
it is easy to imagine that the pairs of nearest spins, 
which is along the bond with hopping $t_{xy}$, 
form essentially a localized singlet. 

There exists a different proposal for the values of the hopping integrals 
with very small $t_{xy}$, 
i.e. $t_a=0.38$ eV, $t_b=0.17$ eV and $t_{xy}\ll 0.1$ eV.\cite{Smolinski}
We note, however, the result of our Hartree calculations 
even for $t_{xy}=0$ eV does not show qualitative difference
from those given in Figs. \ref{energy} and \ref{Sz}. 
We also note that the results are not largely affected by 
the condition $V_a=V_b=V_{xy}/\sqrt{2}$ that we adopted; 
the results for $V_a=V_b=V_{xy}$ show the same features. 

Now we compare our present results of the ground state 
with the `zigzag' pattern of charge ordering, 
and the various experimental results noted above. 
The spin gap behavior in the magnetic susceptibility 
measurement\cite{Isobe} can 
be understood as the spin singlet formation as discussed above. 
This is not a SP state and then 
the comparison of $\Delta/T_{c}$\cite{Fujii} to other SP compounds 
is irrelevant. 
The change in valences through $T_c$ 
observed in the NMR measurement\cite{Ohama}
may be due to the phase transition 
from the `dimer' state above $T_c$ to the `zigzag' state below $T_c$.
The `zigzag' type charge ordering result in 
the period of two V sites in the $b$-direction 
and of four ladders in the $a$-direction (see Fig. \ref{AF}(c)), 
namely doubles the original unit cell size in the $a$-$b$ plane, 
which is consistent with the observed results below T$_c$.\cite{Fujii} 
The exitation spectra from the `zigzag' singlet ground state 
is an unanswered question, 
but a deviation from 1D character is expected. 
Thus it is natural that the inelastic neutron scattering measurement showed  
a dispersion which is not expected from 1D SP chains.\cite{Yoshihama} 
However, more experimetal studies, 
especially the firm determination of crystal structure 
both below and above $T_c$ 
are needed for further discussions, 
e.g. regarding the unit cell size enlargement
along the $c$-direction\cite{Fujii} 
and the intensity modulation along the $a^*$-axis 
in the inelastic neutron scattering measurement.\cite{Yoshihama,comment} 

In summary, 
the ground state of NaV$_2$O$_5$ is investigated 
within the Hartree approximation 
for both on-site and intersite Coulomb interactions. 
The effect of intersite Coulomb interaction results in 
a zigzag pattern of V$^{4+}$ along the ladders of vanadiums
resulting in the singlet pairs,  
which is different from the conventional picture 
that this compound is 
a spin-Peierls system. 
This proposed ground state is consistent with 
the observed spin gap 
and also with the other experimental results 
so far reported. 

\acknowledgements
The authors thank N. Katoh, M. Saito and H. Kohno
for useful discussions suggestions. 
They also thank H. Yasuoka 
for informative discussions from experimental point of view.  
This work was financially supported by a 
Grant-in-Aid for Scientific Research on Priority Area ``Anomalous Metallic
State near the Mott Transition'' (07237102) from the Ministry of Education, 
Science, Sports and Culture.



\begin{thebibliography}{99}
\bibitem{comment}In ref. 4,
it is discussed that this 
unexpected intensity modulation can be explained if the positions of the 
V$^{4+}$ and V$^{5+}$ chains alternate from one layer to another. 
Similarily, the alternation of the pattern of charge ordering 
in the $c$-direction 
in the presence of the `zigzag' pattern may explain this modulation.
\end{thebibliography}
\end{document}